\documentclass[lettersize,journal]{IEEEtran}
\usepackage{amsmath,amsfonts}
\usepackage{algorithmic}
\usepackage{algorithm}
\usepackage{array}
\usepackage[caption=false,font=normalsize,labelfont=sf,textfont=sf]{subfig}
\usepackage{textcomp}
\usepackage{stfloats}
\usepackage{url}
\usepackage{verbatim}
\usepackage{graphicx}
\usepackage{cite}
\everymath{\displaystyle}
\usepackage{multirow}
\usepackage[table]{xcolor}
\usepackage{epstopdf}

\begin{document}

\title{{Performance Analysis of ML-based MTC Traffic Pattern Predictors
}}

\author{{David E. Ruíz-Guirola, \IEEEmembership{Student Member, IEEE}, Onel L. A. López, \IEEEmembership{Member, IEEE}, Samuel Montejo-Sánchez, \IEEEmembership{Senior Member, IEEE}, Richard Demo Souza, \IEEEmembership{Senior Member, IEEE}, and Mehdi Bennis, \IEEEmembership{Fellow, IEEE}}

\thanks{
David Ruíz-Guirola, Onel López, and Mehdi Bennis are with the Centre for Wireless Communications, University of Oulu, Finland. \{{David.RuizGuirola, Onel.AlcarazLopez, Mehdi.Bennis}@oulu.fi\} 
Samuel Montejo-Sánchez is with the {Programa Institucional de Fomento a la Investigación, Desarrollo e Innovación, Universidad Tecnológica Metropolitana}, Santiago, Chile. \{smontejo@utem.cl\}
Richard Souza is with the Federal University of Santa Catarina, Florianópolis, SC, Brazil. \{richard.demo@ufsc.br\}

This work has been partially supported in Chile by ANID FONDECYT Iniciación No. 11200659 and FONDEQUIP-EQM180180, 
in Brazil by CNPq (402378/2021-0, 305021/2021-4), Print CAPES-UFSC “Automation 4.0”, and RNP/MCTIC (Grant 01245.010604/2020-14), and in Finland by 6Genesis Flagship (Grant no. 318927) and Tekniikan Edistämissäätiön.}}

\maketitle

\begin{abstract}

Prolonging the lifetime of massive machine-type communication (MTC) networks is key to realizing a sustainable digitized society. 
Great energy savings can be achieved by accurately predicting MTC traffic followed by properly designed resource allocation mechanisms. 
{However, selecting the proper MTC traffic predictor is not straightforward and depends on accuracy/complexity trade-offs and the specific MTC applications and network characteristics. Remarkably, the related state-of-the-art literature still lacks such debates.} Herein, we 
{assess the performance of} 
several machine learning (ML) methods to predict 
{Poisson and quasi-periodic MTC traffic} 
in terms of 
accuracy and 
{computational cost.}
Results show that the temporal convolutional network (TCN) outperforms the long-short term memory (LSTM), the gated recurrent units (GRU), and the recurrent neural network (RNN), in that order. For Poisson traffic, the accuracy gap between the predictors is larger than under quasi-periodic traffic. Finally, we show that running a TCN predictor is around three times more costly than other methods, while the training/inference time 
{is} the greatest/least. 

\end{abstract}


%
%
\section{Introduction}
\IEEEPARstart{T}{he} Internet of Things (IoT) {promises} to interconnect everything towards a data-driven society~\cite{WCL_IoT}. A key enabler of IoT is machine-type communication (MTC), where devices exchange information without human intervention~\cite{zhao2022quality}. {MTC devices (MTDs) facilitate} a wide range of applications such as intelligent surveillance, smart agriculture, and autonomous driving~\cite{manta}, {by delivering} data through bandwidth-constrained networks to more specialized devices for further processing. 
The MTDs are usually deployed in large-scale areas and might use limited-capacity batteries that cannot be recharged or replaced. Energy-efficient techniques are necessary for prolonging the network {lifetime and avoiding frequent battery replacement for a massive number of MTDs}
\cite{mehmeti2022modeling}. 



Traffic prediction can {enable efficient network resource scheduling and}  avoid the potential energy waste resulting from idle listening and channel access contention in dense MTC networks~\cite{mehmeti2022modeling}. 
However, the instant in which an alarm {event triggers information exchange} by MTDs is generally unknown and has to be estimated by continuous observations, draining the MTDs' battery.
Understanding the MTC traffic characteristics and designing {proper} traffic predictors is key to realizing energy-efficient MTC networks.
In this regard, machine learning {(ML) algorithms} are very appealing~\cite{wang20206g}.
{
In general, ML-based techniques allow the system to learn from data and thus optimize its overall operation in real-time, increasing resource utilization and energy savings~\cite{zakarya2017energy}.} 

In this line, authors in~\cite{weerasinghe2019supervised} propose a supervised ML model to predict bursty MTC traffic arrivals, and thus avoid collisions and long latency. Long-short-term memory (LSTM) mechanisms are proposed in \cite{weerasinghe2020preamble,preamble,senevirathna2020event,IoT_FWuS,chen2019intelligent,eslam}.  Specifically, the approaches in \cite{weerasinghe2020preamble,preamble} aim to predict bursty MTC traffic and congestion, while event-driven traffic is considered in~\cite{senevirathna2020event,IoT_FWuS}. In the case of \cite{IoT_FWuS}, 
the prediction is used to properly tune wake-up parameters to avoid frequent page monitoring occasions in idle states and promote energy savings. 
Meanwhile, 
the proposal in \cite{chen2019intelligent} seeks to effectively predict the peak traffic flow to ultimately reduce both latency and the packet loss rate, while the authors in \cite{eslam} develop a fast uplink grant for massive MTC (mMTC) by predicting which devices are active at each time instant and then classifying their priorities.
Finally, authors in \cite{shoaei2021traffic} propose a neural network (NN) algorithm that exploits device traffic correlations for enhanced prediction. 

Despite recent advances, adequate performance comparisons between different 
{ML predictors for MTC traffic, especially in terms of accuracy and complexity, still do not exist.
This makes it difficult the selection of appropriate ML models, especially in applications demanding energy efficiency~\cite{mughees2020towards}.
Herein, 
we} take initial steps to fill this gap. For that sake, we overview different approaches such as recurrent NN (RNN), gated recurrent units (GRU), 
LSTM, and temporal convolutional network (TCN) to identify and predict 
{Poisson and quasi-periodic} MTC traffic patterns. 
We analyze and discuss their performance trade-offs in terms of accuracy, memory, and response 
{time. 
The} results evince that the TCN-based predictor provides the best accuracy, although the computational cost is higher compared to other models. Moreover, our approach provides a unified framework for assessing the performance of RNN, GRU, 
LSTM, and TCN under different traffic models.



\section{System model}\label{sec2}
\begin{figure}[t!]
	\centering
	\includegraphics[width=0.90\columnwidth]{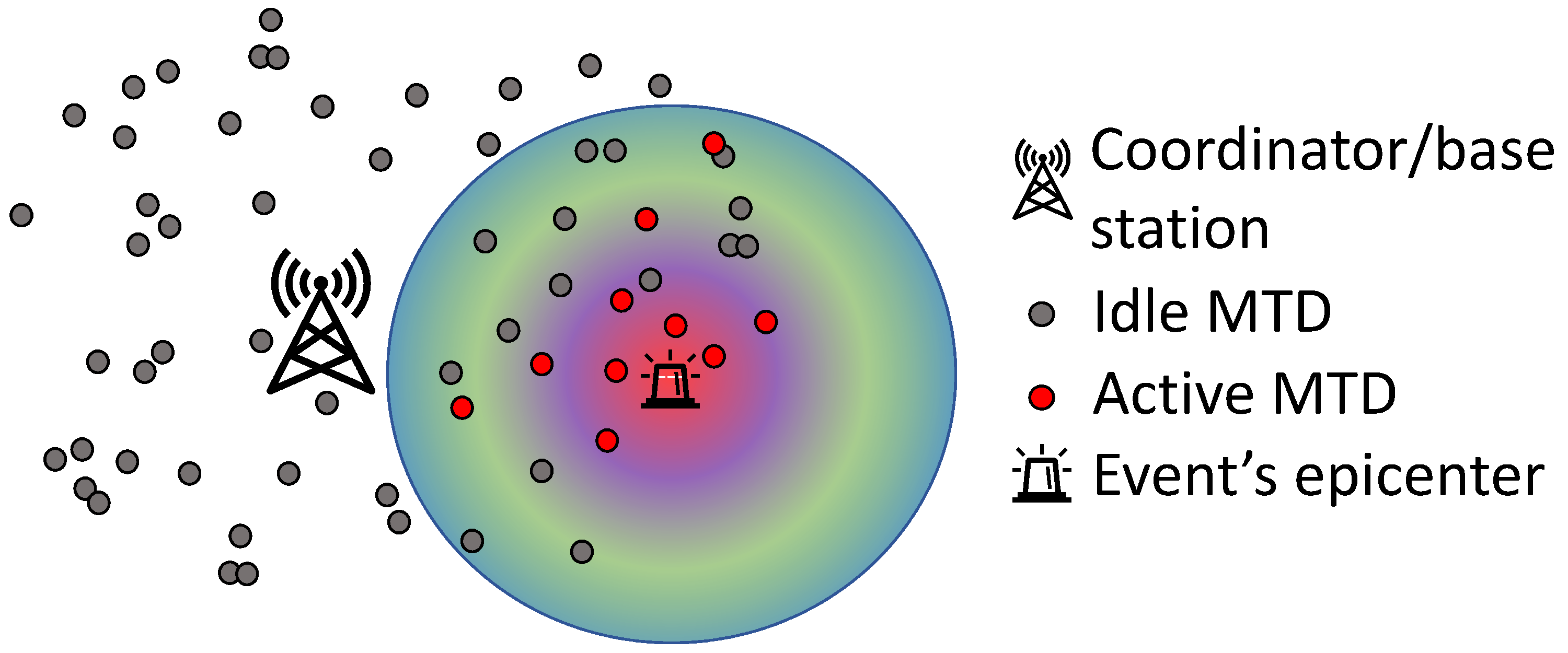}
		\vspace{-3mm}
	\caption{Illustration of an MTC network in which a coordinator controls and collects information from the MTDs. 
	The influence of events on the surrounding MTDs is modeled by a probability function that depends on the distance from the event epicenter to the MTDs.
	}
	\vspace{-4mm}
\label{figure1}
\end{figure}

We consider a single coordinator/base station, which serves as the gateway of short-range MTDs as depicted in Fig.~\ref{figure1}. The MTDs send packets to the coordinator, which controls all the information exchange within its cell. 
Each MTD can be idle (\textit{I}), waiting for a triggering event, or active (\textit{A}), exchanging information with the coordinator. The transition from state \textit{I} to \textit{A} occurs when 
information exchange between the MTD and coordinator is triggered due to the detection of an event. 
When the MTD goes to state \textit{A}, it stays there for the duration of the event. 
Assume that time is slotted in transmission time intervals (TTI). In time slot \textit{k} and state 
\textit{A} the {MTD}s  generate traffic with rate \textit{R}. In state \textit{I} the MTDs do not generate traffic.

To model the position of MTDs and event epicenters, we use Poisson point processes (PPPs) as nodes and events can be assumed to be stochastically deployed in the Euclidean plane~\cite{IoT_FWuS}. The MTDs are deployed according to a 2D homogeneous PPP $\Phi_{M}$ with density $\lambda_{M}$. The event epicenters are represented by a 2D homogeneous PPP $\Phi_{E}$ with density $\lambda_{E}$. The processes $\Phi_{M}$ and $\Phi_{E}$ are assumed to be independent, while the coordinator is 
at the origin, as in Fig.~\ref{figure1}.
\subsection{Influence of an Event Epicenter}\label{sec22}
To capture the effect of a given event on a sensing MTD, we define a function $p(d_{i,j})$ as the probability that an event in the $i^{th}$ epicenter ($i\in\Phi_E$) triggers an MTD $j$ at a location $j\in\Phi_M$, where $d_{i,j}$ is the distance between them in the Euclidean plane $\Re^{2}$~\cite{thomsen2017traffic}. Moreover, $p(d_{i,j}): [0, \infty) \rightarrow [0,1]$, 
is non-increasing to mimic a decaying influence of events as the distance $d_{i,j}$ increases~\cite{IoT_FWuS}. Fig.~\ref{figure1} depicts 
the influence of an event epicenter on the surrounding MTDs.
\section{MTC traffic models}\label{sec21}
The MTC traffic is usually uplink-dominated and characterized by short transmissions combining real-time and non real-time traffic from multiple sources. According to its applications, MTC has three elementary traffic patterns~\cite{lopez2021csi}: (i) periodic update (PU), under which {devices transmit status reports 
regularly, \textit{e.g.}, smart meter reading (gas, electricity, water)}; (ii) event-driven (ED), which describes non-periodic traffic due to a 
specific random trigger at an unknown time, \textit{e.g.}, alarms; and (iii) payload exchange (PE), which consists of {bursty traffic that usually comes after PU or ED traffic.} 

The MTC traffic is often a combination of the aforementioned  types~\cite{IoT_FWuS}. For instance, an MTD may enter the power saving mode and trigger a PU pattern at regular intervals, while an alarm or critical event may activate the MTD and originate ED followed by PE traffic. Hence, using the three elementary classes above enables building traffic models with an arbitrary degree of computational complexity and accuracy~\cite{eslam}.
\subsection{Generation of Events}\label{sec23}

The events are generated over time according to:

\subsubsection{Poisson model}\label{sec231}

The 
time between the occurrence 
of 
events 
follows a Poisson distribution with density $\lambda_{T}$
. Then, 
each MTD goes to state \textit{A} with probability
\begin{equation}
	P_{A} = 1 - \exp\left({-2\pi\lambda_{T}\int^{\infty}_{0}{p(d)\partial d}}\right). 
    \label{PPPtraffic}	
\end{equation}

\subsubsection{Quasi-periodic traffic}\label{sec232}

This pattern is typical in industrial IoT~\cite{mitev2022smart}. 
Here, we consider MTC traffic characterized by homogeneous asynchronous periodicity. The transfer intervals $T_j$ for the $j^{th}$ MTD, the number of time slots between consecutive transmissions, are independent and quasi-identically distributed~\cite{mitev2022smart}. The coordinator receives the MTDs' signals at independent start times ($\kappa_j$). The activation probability and transmission duration for each MTD are denoted by $P_{A_j} \in [0,1]$ and $\delta_j$ respectively. The latter denotes the number of time slots required for each transmission. 
Finally, 
packets from a single MTD are transmitted 
with start time $t_{j}$ such that
\begin{equation}
    t_j = (\kappa_j + (m-1)T_j)B_j, \text{ } m = 1, 2, \dots ,
    \label{quasiperiodic}
\end{equation}
where $B_j$ is a Bernoulli random variable with parameter $P_{A_j}$, and $m$ denotes 
the transmission opportunity. 
\subsection{Payload exchange}\label{sec24}
The PE patterns, whose durations are quantified by $\delta_j$, are modeled through the 
geometric distribution. The parameter $q$ of the geometric distribution tunes the burstiness of the traffic generated by an event. Specifically, once in state \textit{A}, the MTD remains there for a number $k$ of TTIs with probability 
\begin{equation}
	{G_{k}} (q)= (1-q)q^{k}, \ k = 0,1,\dots
\label{geometric}	
\end{equation}
We assume $T_j$ to be large enough so that the probability 
that 
two transmission opportunities overlap due to a relatively large $\delta_j$ ($k$ TTIs) is almost zero. 
Note that  
 parameter $q$
allows tuning the temporal correlation of the individual rate processes of the MTDs and that of the total rate process, mimicking various MTC applications. For instance, in the case of small $q$, the traffic behaves similarly as a Bernoulli process (memoryless)~\cite{IoT_FWuS}. As $q$ increases, so does the memory since the total rate at a given time $k$ is correlated with many past values. Then, once state \textit{A} 
is entered, one stays there longer~\cite{IoT_FWuS}. The traffic exchanged between the coordinator and the MTDs, when following a Poisson model, may be modeled using an ergodic Markov chain~\cite{IoT_FWuS} with two states, \textit{I} and \textit{A}.

\begin{figure}[t!]
	\centering
	\centerline{\includegraphics[width=0.99\columnwidth]{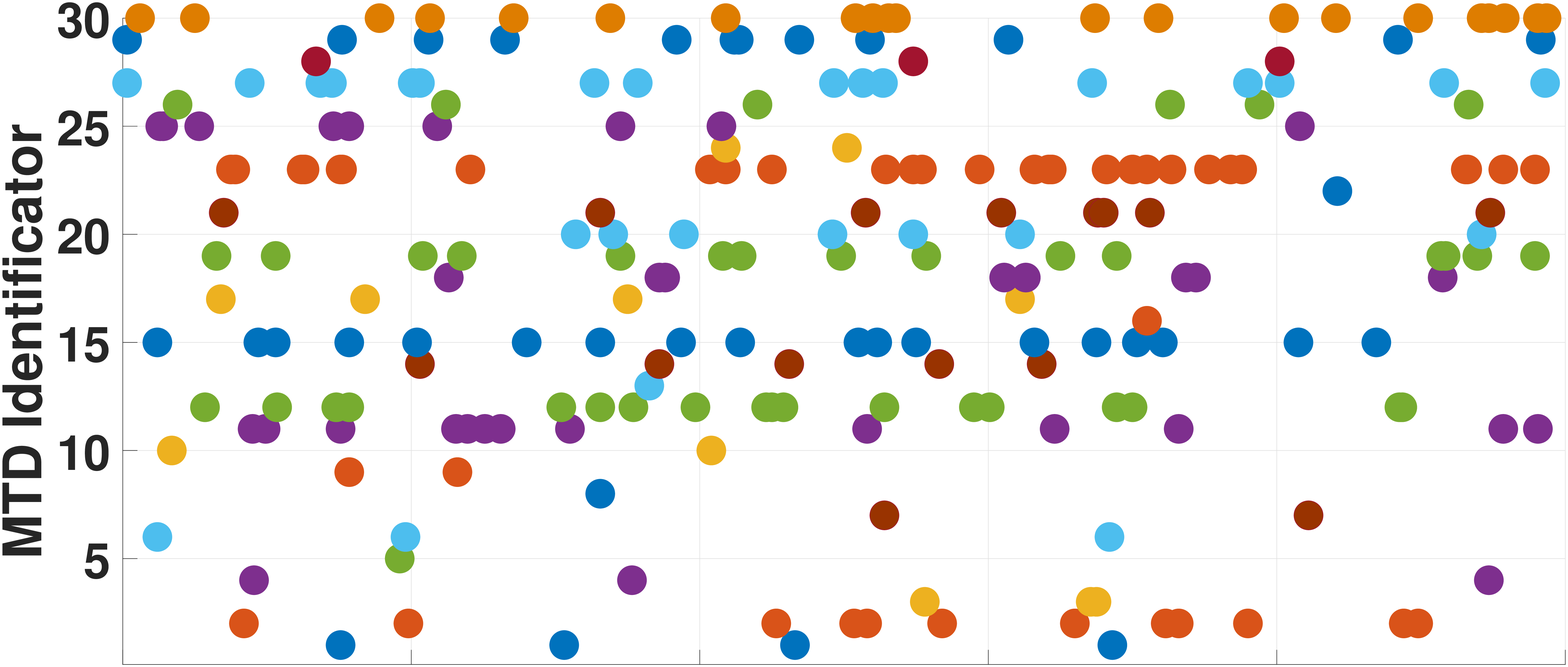}}
	\vspace{-10mm}
	
	\centerline{\includegraphics[width=\columnwidth]{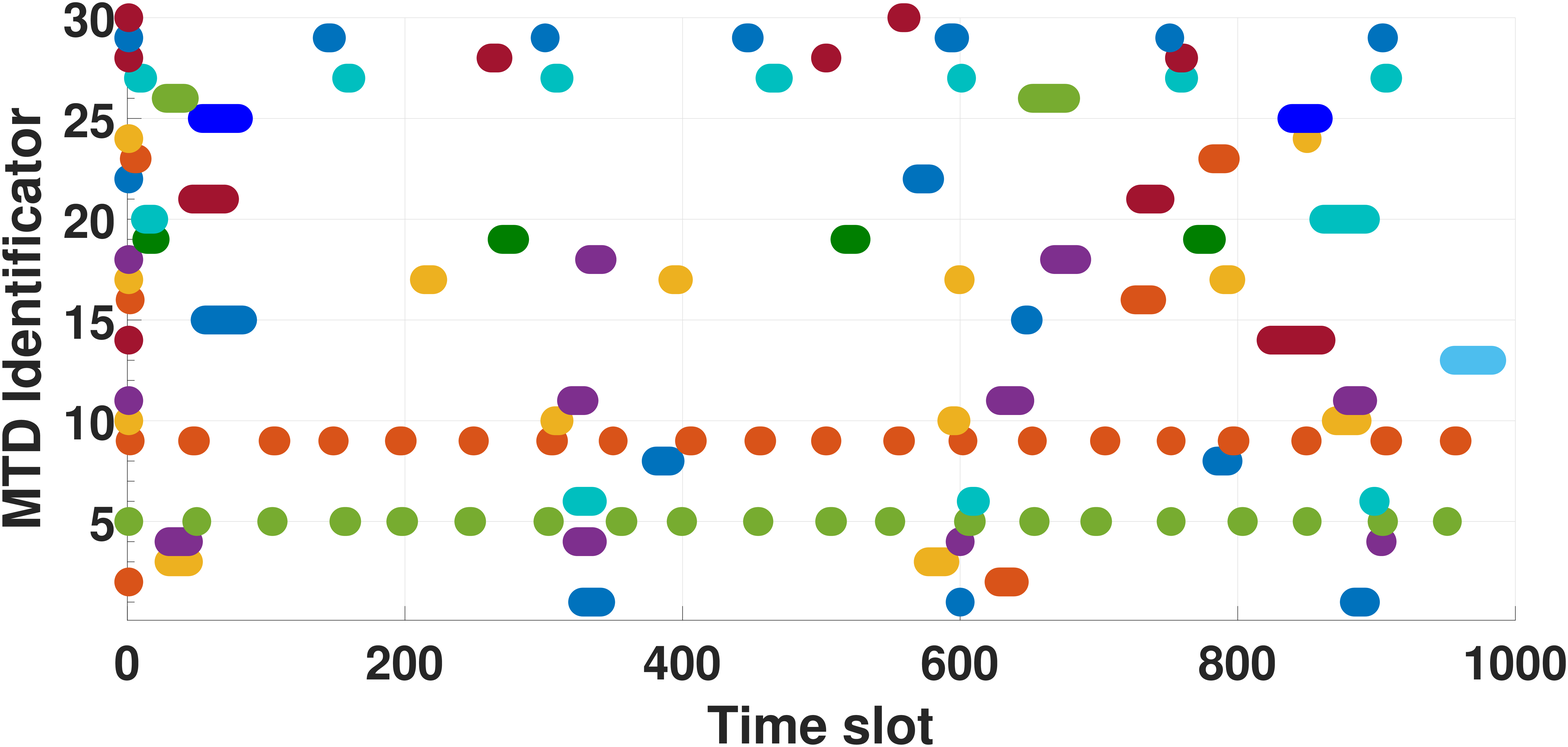}}
	\vspace{-4mm}
	\caption{Extract of a) Poisson (top), and b) quasi-periodic (bottom) traffic. The dots represent the time slots (x-axis) where a given MTD (y-axis) is Active.}
	\vspace{-4mm}
\label{quasi_traffic}
\end{figure}
Fig.~\ref{quasi_traffic} shows extracts from the MTDs traffic. Due to unequal activation probabilities, some MTDs have periodic traffic patterns, while others are rarely active.
{Specifically, the values in the y-axis represent 30 randomly selected MTDs, while the x-axis is a given time frame of $10^3$ ms extracted from the traffic. Moreover, the dots represent the time slot in which the corresponding MTD in the y-axis generates data packets. The bottom figure shows the quasi-periodic behavior of the traffic pattern, while the top figure shows a Poisson-like behavior.}
{These models fit accurately real MTC traffic, such as the traffic associated with the measurements of temperature, light, CO$_2$, sound, and humidity from the Smart Campus network at the University of Oulu \cite{realdata}\footnote{{Specifically, the models characterize the real data with an estimation error of less than 9\% and a root mean square error (RMSE) of 0.947 for event-driven traffic and 0.982 for quasi-periodic traffic.}}.}
\section{Traffic Predictors}\label{sec3}
{Herein, we overview the ML predictors used in this paper. 
Such ML mechanisms are deployed at the coordinator side to keep the MTDs simple and energy-efficient. The timestamp of the traffic data received by the coordinator is used as input for the ML mechanisms to train the forecasting model, which is then used to optimize the network performance\footnote{{Although we assume the learning occurs in the coordinator, the model can be trained offline
, while retraining just in case of prediction deviation. Moreover, the training can be performed using cloud computing and just updating the trained model at the coordinator in scenarios where both the MTD and the coordinator should be kept simple and energy efficient.}}. 
} {The MTDs produce explosively large amounts of data. However, what is critical to the ML servers is the characteristics of the data rather than the data itself~\cite{hu2021distributed}. 
Precisely, to predict the arrival time of the next packet, the coordinator only needs the trained model and the history of the previous packet timestamps.
}

\subsection{Recurrent Neural Network (RNN)}\label{sec31}
NNs are flexible nonlinear models compounded of units (neurons) that learn patterns from data. Given an appropriate number of nonlinear processing units, NNs can learn from experience and estimate any complex functional relationship with high accuracy~\cite{duc2020convolutional}. 
Moreover, RNNs constitute powerful NN dynamic systems for modeling long-term dependencies in sequential data
~\cite{duc2020convolutional}. The RNN takes one element in the sequence at each time step and merges the current input with  information from the past time steps to learn inherent patterns. Traditional RNNs suffer from the gradient vanishing problem during training due to recurrent operations. The difficulties in training traditional RNNs are addressed by 
modern 
RNN variants, such as LSTM and GRUs.

    {{\textit{1) LSTM}}\label{sec312}
is a type of RNN where predictions are made based on long sequences of previous input values rather than on a single value~\cite{xu2021generative}. 
LSTM-based techniques can improve significantly the learning speed, especially in problems with large state/action spaces. An LSTM cell is made up of three gates: the input, the output, and a forget gate. These gates determine if the information is read (input gate), if it is not relevant and is disregarded (forget gate), or if it is saved, impacting the current time step (output gate).}
    
    {{\textit{2) GRU}}\label{sec311}
comprises only two gates, namely, an update and a reset gate. GRU is similar to LSTM, but saving one gating signal and the associated parameters for training~\cite{gru}.}

In this work, we build up a traditional RNN, an LSTM, and a GRU architecture with $h$ hidden layers with $n_1, n_2, \cdots,$ and $n_h$ neurons, initial learning rate ($\epsilon$), root mean square error (RMSE) loss function, maximum number of epochs ($n_{ep}$)
, and using the Adam optimizer~\cite{sharma2017adam}. RNN's learning algorithm is local in space and time; while its computational complexity per time step and weight is $ \mathcal{O}(1)$~\cite{gers2000learning}, thus, leading to a computation complexity $\mathcal{O}(2\times n_1 \times n_2 \times \dots \times n_h)$.
\subsection{Temporal Convolutional Network (TCN)}\label{sec32}
A TCN architecture uses a causal convolutional layer to ensure there is no information leakage from future to past and to capture complex dependencies in sequential data. This layer allows the network to operate on a larger scale than a regular convolutional NN~\cite{duc2020convolutional}. 
\textcolor{black}{We build up a TCN architecture with $l$ convolutional layers, 
dropout regularization for reducing overfitting (nodes dropout factor $\rho$), number of filters ($\omega$),  neurons per layer that connect to the same region (receptive field, $r$), 
convolutional filter ($f$) of size $1 \times z$, exponential dilation ($\Lambda$) equal to $2^{i}$ for layer $i$, 
initial learning rate ($\epsilon$), using the Adam optimizer~\cite{sharma2017adam}. The complexity is $\mathcal{O}(f\times \omega \times r)$.} 

\textcolor{black}{The causal convolution ensures that given a series input $x_0, x_1, \dots, x_N$, the output at a time $t$ depends only on the inputs at $t$ and earlier, i.e., $x_t, x_{t-1}, \dots, x_{t-r+1}$ for a given $r$. Meanwhile, the dilated causal convolution, i.e.,
${D{(t)}} = \textstyle\sum_{i=0}^{z-1}f(i)x_{t-{\Lambda i}}$, 
enables reaching a large $r$ with few layers since ($\Lambda$) increases exponentially with every layer. 
However, the local connection among adjacent time steps is not fully extracted at higher layers due to the dilated causal convolution~\cite{duc2020convolutional}. 
One disadvantage of TCNs compared to RNNs is the large memory footprint during inference since the entire sequence must be computed in the next time step.}


\vspace{-4mm}
\section{Performance evaluation and discussions}\label{sec4}
The traffic models in Section~\ref{sec2} are used to simulate several traffic traces  at the coordinator. In each trace, the MTDs 
and 
the event epicenters are randomly deployed 
in the Euclidean plane according to $\Phi_M$ and $\Phi_E$, respectively. The events are independent, while we assume a negative exponential function to model the influence of events on 
each MTD traffic as 
$p(d)=\exp{(-d)}$. Then, $\textstyle \int_0^{\infty}p(d) = 1$ and 
$P_A = 1 - \exp{(-2\pi\lambda_T)}$.

Two traffic models are used: Poisson (Section~\ref{sec231}) with $\lambda_T \in [1,50]$, and quasi-periodic (Section~\ref{sec232}). In the case of the latter,
the transfer intervals for each MTD are in the range of 50 $-$ 1000 ms (with variation upper bounded by $\pm$ 5\% and $\pm$ 10\%) and the time required for transmission follows a geometric distribution as in Section~\ref{sec24}. The traffic data are then used to train the forecasting algorithms. Out of 9.8$\times 10^4$  data samples, 70$\%$ are used to perform training, 15\% for the validation set, and the remaining 15\% for testing. The MTDs are deployed with density $\lambda_{M} = 10^{-1} (\text{MTDs}/m^{2})$, and the TTI is assumed equal to 1 ms. All MTDs are at the state \textit{I} at the beginning, and 
$R = 1$ packet/TTI. These parameters, summarized in Table~\ref{tab2}, are used for all simulations.  

The performance of each predictor is the result of 150 Monte Carlo runs, where the position of the MTDs and the events’ epicenter are randomly distributed in each run.
{As a comparison baseline, we use autoregressive integrated moving average (ARIMA), a well-known linear statistical model proposed by Box and Jenkins~\cite{box2015time} for time series prediction.} 
For fairness purposes, the configuration of the predictors was conceived so that all share the same big-$\mathcal{O}$ complexity. 

\begin{table}[t!]
    \caption{Simulation Parameters}
    \vspace{-2mm}
    \centering
    \label{tab2}
    \begin{tabular}{lll}
        \hline
        Parameter    &Value   &Reference\\
        \hline
        $\rho$ (dropout factor)
        &0.05   &\cite{
        sadique2022modeling}\\
        $z$ (filter size)
        &2    &\cite{duc2020convolutional, 
        sadique2022modeling}\\
        $l$ (convolutional layers)
        &3      &\cite{duc2020convolutional, 
        sadique2022modeling}\\
        $n$ for RNNs (hidden neurons)
        &256   &\cite{IoT_FWuS, eslam}\\
        $n_{ep}$ (epochs)
        &50     &\cite{eslam,IoT_FWuS}\\
        $\omega$ (number of filters)
        &32   &\cite{duc2020convolutional
        }\\
        $r$ (receptive field)
        &8      &\cite{duc2020convolutional,
        sadique2022modeling}\\
        $\epsilon$  (initial learning rate)
        &$10^{-3}$      
        &\cite{sadique2022modeling}\\
        \hline
    \end{tabular}
    \label{tab2}
\end{table}

\begin{table}[t!]
    \caption{Performance Comparison of the Traffic Predictors for Poisson (gray) and quasi-periodic (white) traffic patterns}
    \vspace{-2mm}
    \centering
    \label{quality_metric}
    \begin{tabular}{lccccc}
        \hline
        Performance indicator      &\textbf{RNN}            &\textbf{LSTM}              &\textbf{GRU} 	          &\textbf{TCN} &{\textbf{ARIMA}}
        \\ 
        \hline
        \rowcolor{lightgray}
        R metric   	    &0.928              &\textbf{0.962}    &0.941      	&0.959	&{0.894}
        \\
                        &0.968     &0.989		        &0.981	    	  &\textbf{0.991}  &{0.953}
                        \\
        \rowcolor{lightgray}
        TPR   		    &0.896     &0.912             &0.908		    &\textbf{0.937}   &{0.852}
        \\
                        &0.951     &0.981		        &0.969		&\textbf{0.986}	    &{0.947}
                        \\
        \rowcolor{lightgray}
        TNR   		    &0.931     &0.987             &0.965	    &\textbf{0.991}	    &{0.926}
        \\
                        &0.976    &\textbf{0.995}		&0.986	    &\textbf{0.995} 		&{0.964}
                        \\
        \rowcolor{lightgray}
        Accuracy   	    &0.914     &0.949             &0.937		&\textbf{0.964}		&{0.889}
        \\
                        &0.964     &0.988		        &0.978		&\textbf{0.990}		&{0.958}
                        \\
        \hline
    \end{tabular}
    \label{quality_metric}
    \vspace{-4mm}
\end{table}
\subsection{Numerical Results}\label{sec41}
Table~\ref{quality_metric} compares the predictors in terms of recall (R), true positive rate (TPR), true negative rate (TNR), and accuracy. 
{Specifically, R is an absolute measure of the ability to perform accurate predictions and is given by 
$\textstyle{\text{R} = \big(1-\frac{1}{M}\sum_{i=1}^{M}{{(1-{\hat{g}_i}}/{g_{i})^{2}}}}\big)^{1/2},$ where $\hat{g}_i$ and $g_i$ are the estimated and actual inter-arrival values and $M$ represents the sample size. The higher {R metric}, the better the data fits. TPR, computed as true positive/(true positive + false negative),} gives a measure of the false alarm probability\footnote{A higher TPR (lower false alarm probability) implies a better energy efficiency, as communication attempts with inactive MTD are avoided.}. 
{TNR, computed as true negative/(true negative + false positive),} 
measures the miss-detected information. Meanwhile, 
{the accuracy metric constitutes} a mix of TPR and TNR 
{and is calculated as (number of correct predictions)/(total number of predictions).}
{{TCN mostly outperforms the other, {while ARIMA performs worst.}} Moreover, while the gap from TCN to the others is not significant for quasi-periodic traffic, for Poisson traffic  LSTM  is the only one that comes close to  TCN.} 

\begin{figure}[t!]
    \centering
    \centerline{\includegraphics[width=\columnwidth]{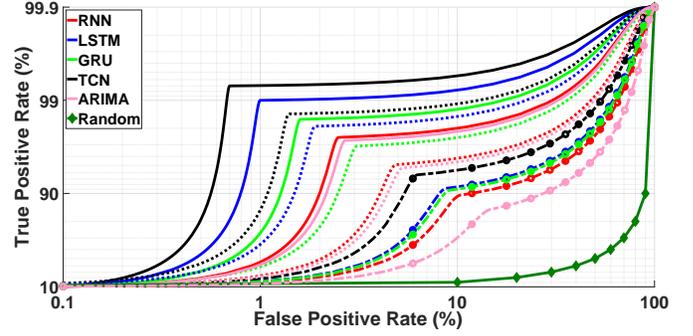}}
    \vspace{-2mm}
    \caption{{ROC curves related to each model, for the quasi-periodic traffic with $\pm 5\%$ variation (straight lines) and $\pm 10\%$ variation (dotted lines), the Poisson traffic model (dash-dotted lines with circle marker). Note that the performance of the purely random method does not depend on the traffic model used
    .}}
    \label{ROC}
    \vspace{-4mm}
\end{figure}

Fig.~\ref{ROC} illustrates the receiver operating characteristic (ROC) curves for both traffic models. Notice that TCN outperforms the other predictors, although not significantly with respect to LSTM. {Observe that the false alarm probability (1 $-$ TPR) is around 1\% for LSTM and TCN, while for the others it is up to 7.4\%.}  
This metric is relevant when modeling scenarios sensitive to delay and packet loss. 
Fig.~\ref{ROC} shows the robustness of the predictors to data distribution changes, where TCN and LSTM are the most adaptable against variations in the traffic pattern. 
Note that LSTM with quasi-periodic traffic variations of 10\% outperforms RNN 
with 5\%, and TCN with 10\% outperforms GRU with 5\% traffic variations. 
For Poisson traffic, it is noteworthy that TCN outperforms the other predictors while LSTM and GRU have similar performance.

It is noteworthy that the predictors 
perform relatively well under 
quasi-periodic traffic conditions, even with variations up to almost $\pm$30\% in TCN and LSTM cases, and up to $\pm$25\% in the rest. Meanwhile, under Poisson traffic, the prediction accuracy 
is more seriously affected due to the inherently stronger randomness. 
However, slight variations in Poisson traffic have no significant impact on the predictors' accuracy.
\subsection{Complexity Analysis}\label{sec42}
\textcolor{black}{Table~\ref{complexity} shows the performance in terms of inference time, training time, and model size. 
The three parameters are normalized relative to the 
total considering the contribution 
of each architecture. 
Notice that 
TCN and RNN have respectively the largest and smallest model size. 
Meanwhile, regarding inference time, TCN and LSTM are the fastest and lowest, respectively. However, LSTM needs less training time to reach a viable forecasting model while TCN requires more.} 

{In any case, a deeper network architecture would enhance the prediction accuracy by enabling the extraction of more relevant information, but at the expense of an increased complexity, training and inference time. 
Notice that 
model size and 
training data size should be adjusted according to the hardware at the BS, the requirements of the application, the number of devices, and 
target packet error rate~\cite{eslam}.}

\begin{table}[t!]
    \caption{Complexity for each ML architecture}
    \vspace{-2mm}
    \centering
    \label{complexity}
    \begin{tabular}{lcccc}
        \hline
            &\textbf{RNN}   &\textbf{LSTM}   &\textbf{GRU}   &\textbf{TCN}\\
        \hline
        \textbf{Inference time}  &{0.222}    &0.324    &{0.271} &\textbf{0.183}\\
        \textbf{Training time}   &0.304    &\textbf{0.145}  &0.213  &0.338\\         
        \textbf{Model size}  &\textbf{0.167}    &0.229    &0.181     &0.423\\
        \hline
    \end{tabular}
    \label{complexity}
    \vspace{-2mm}
\end{table}

\begin{figure}[t!]
    \centering
    \centerline{\includegraphics[width=0.95\columnwidth]{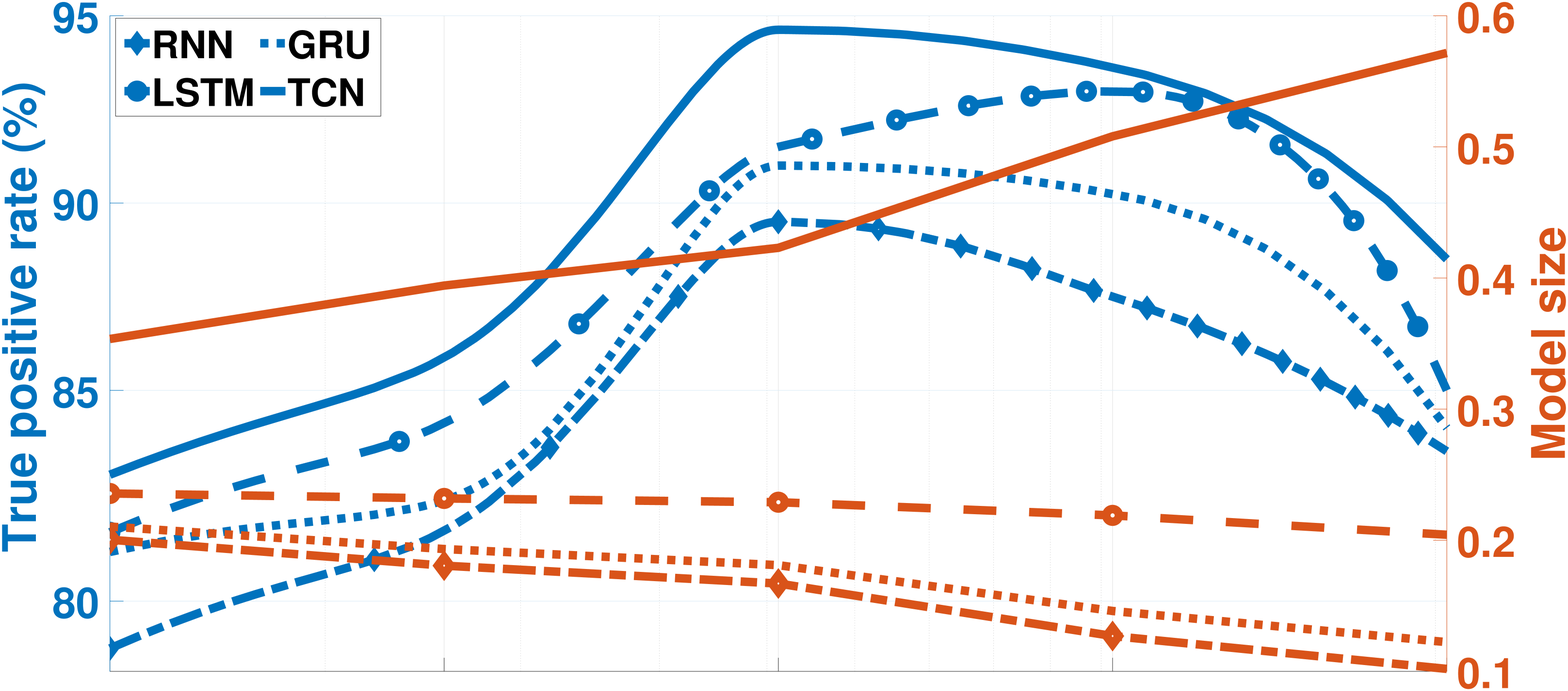}}
    \vspace{-6mm}
    \centerline{\includegraphics[width=0.95\columnwidth]{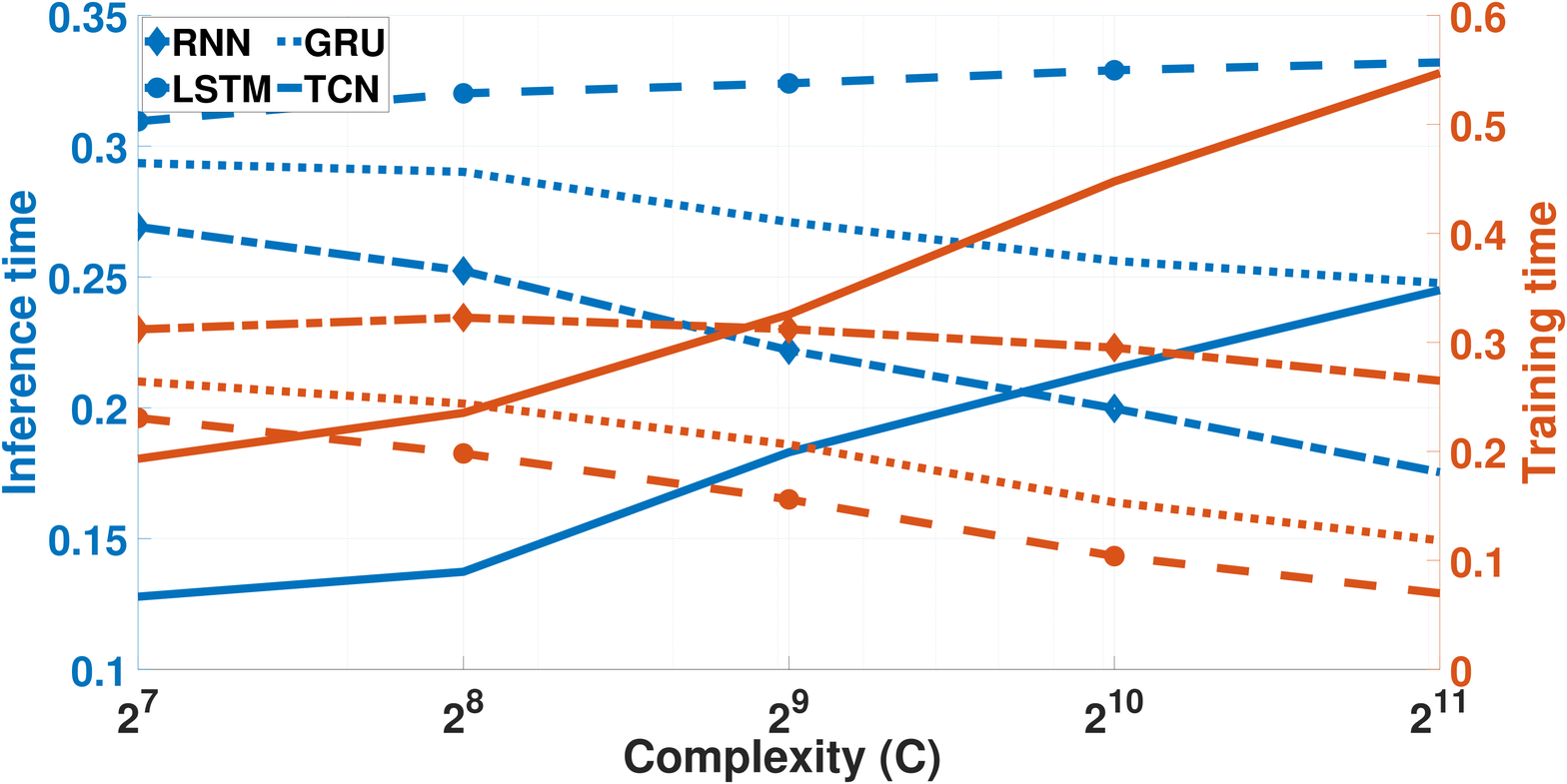}}
    \vspace{-2mm}
    \caption{{Performance in terms of a) true positive rate and model size (top), and b) inference and training time (bottom), as a function of complexity $\mathcal{O}$(C) and for Poisson traffic. Values in y-axes in b) and right y-axis in a) are normalized.}}
    \label{cost}
    \vspace{-5mm}
\end{figure}

Fig.~\ref{cost}a) shows the performance versus model complexity. TPR increases when a deeper architecture is used up to complexity (C) below 1024, while above that the models tend to overfit. The gap between TCN and RNN decreases after reaching the best performance. LSTM, which attains the second best performance, needs a deeper architecture to reach its best prediction performance. GRU converges faster than LSTM, thus reaching the best performance with a less complex model. This is because 
GRU typically
has fewer trainable parameters~\cite{sadique2022modeling}. All in all, TCN shows the best prediction accuracy 
regardless of the network depth. Regarding model size, a deeper TCN architecture demands more computational resources (right y-axis), increasing the gap to other models, while the gap between LSTM and GRU increases slightly. 

Fig.~\ref{cost}b) shows the relative time for different model complexities.  
For inference time (left y-axis), LSTM is the slowest regardless of the complexity, while the gap between TCN and GRU decreases with the complexity. \textcolor{black}{For training time, a deeper architecture increases the gap between TCN and the others, while LSTM has the fastest convergence.}  

\vspace{-3mm}
\section{Conclusions}\label{sec5} 
We compared and analyzed several 
traffic forecasting methods. 
We considered a system model using independent Poisson point processes for spatial modeling of MTDs and event epicenters. Furthermore, Poisson and quasi-periodic traffic patterns were modeled while taking into account event-driven traffic patterns with geometrically distributed burst duration. We showed the superiority of TCN and the extremely poor accuracy attained by RNN. 
\textcolor{black}{Under quasi-periodic traffic, LSTM and TCN outperform the other methods with similar results in terms of prediction accuracy. However, the former is superior in terms of inference time, while the training time is smaller when using LSTM.} On the other hand, the cost of running a TCN-based predictor is far higher (around 3 times) than the other baselines in terms of memory footprint. 

\bibliographystyle{IEEEtran}
\bibliography{bib.bib}

\vfill

\end{document}